# Handling Overload Conditions In High Performance Trustworthy Information Retrieval Systems

Sumalatha Ramachandran, Sharon Joseph, Sujaya Paulraj, Vetriselvi Ramaraj

**Abstract**– Web search engines retrieve a vast amount of information for a given search query. But the user needs only trustworthy and high-quality information from this vast retrieved data. The response time of the search engine must be a minimum value in order to satisfy the user. An optimum level of response time should be maintained even when the system is overloaded. This paper proposes an optimal Load Shedding algorithm which is used to handle overload conditions in real-time data stream applications and is adapted to the Information Retrieval System of a web search engine. Experiment results show that the proposed algorithm enables a web search engine to provide trustworthy search results to the user within an optimum response time, even during overload conditions.

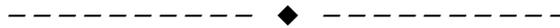

## 1 INTRODUCTION

THE increasing amount of information available in the web reduces the guarantee of information correctness. Conflicting information of varying quality are provided by diverse sources of information on the web. Hence trustworthy and high-quality search results must be given to the user [1]. But when the search engine is overloaded with a large amount of retrieved search results, it becomes difficult to process all the URLs to calculate their trustworthiness [1] within an optimum response time. The concept of Load Shedding in real-time Data Stream Management Systems (DSMS) is used to reduce the response time of search engines during overload conditions.

A data stream is a real-time, continuous, ordered sequence of data items [2]. The main characteristic of a DSMS is that, data tuples are processed as they enter the system without storing them explicitly [5]. Hence the use of DSMS reduces the memory requirements of the system. Since the size of the data streams is potentially unbounded, there is a possibility for a real-time data stream system to be overloaded. Load Shedding in DSMS helps in reducing the Deadline Miss Ratio (DMR) [2] even when the system is overloaded.

The URLs retrieved in the search results can be considered as the real-time data stream and the optimum response time to be achieved by the web search engine is taken as the deadline of the real-time system under consideration.

This paper proposes an optimal Load Shedding algorithm which handles overload conditions in the system and also meets the deadline of the system i.e., the optimum response time.

The rest of the paper is organized as follows. Section 2 briefly explains the related works. Section 3 provides the problem definition and the proposed work. Section 4 explains the architecture of the Load Shedding module. Section 5 explains the optimal Load Shedding algorithm. Sections 6 and 7 provide the analysis of final results and conclude the paper respectively.

## 2 RELATED WORKS

A framework for retrieving trustworthy and high-quality information from the web is given in [1]. The framework contains two modules, the Trust Module and Quality Module. The trust module aims at calculating the trustworthiness of the URLs retrieved in the search results, before displaying them to the user. This processing must be performed for all the URLs for an accurate trustworthiness calculation.

When the number of search results exceeds a certain threshold level, system overload occurs and the trust module is unable to provide the search results to the user with minimum response time value. As the number of URLs increases, the time taken to process all the URLs also increases

————————————————
- *Sumalatha Ramachandran is with the Anna University, Chennai, India. E-mail: rsumalatha@rediffmail.com*
- *Sharon Joseph is a student of the Department of Information Technology, Anna University, and Chennai, India.*
- *Sujaya Paulraj is a student of the Department of Information Technology, Anna University, and Chennai, India.*
- *Vetriselvi Ramaraj is a student of the Department of Information Technology, Anna University, and Chennai, India.*

tremendously. Hence the framework proposed in [1] cannot be considered as efficient during overload conditions.

Thus, a new framework which handles the overload conditions of the system as well as providing trustworthy search results within an optimum response time is needed.

In [3], [4], [8], a feedback control technique is used to handle load shedding problem in DSMS. In [7], A Semantic method is used in discarding the extra load in DSMS. In [2], Effective Deadline-Aware Random Load Shedding algorithm (RLS-EDA) is proposed, which provides an effective solution for Load Shedding problem in DSMS. When the data stream overloads a system load shedding approach is used to handle the situation. The limitation in this algorithm is that after a threshold limit is exceeded the algorithm sheds the extra load without processing them. This reduces the system's ability to provide an accurate result. The algorithm can process the stream of data within deadline but cannot provide an accurate result.

## 3 PROPOSED WORK

To provide trustworthy search results to the users in short time, the response time of the search engine must be reduced significantly. Hence the concept of DSMS is used to reduce the response time of the search engine. The resultant URLs of the search query is the data stream which is processed (Calculation of trustworthiness) without explicit storage. The deadline of the system is taken as an optimized response time which the system aims to achieve.

Limitation of the RLS-EDA algorithm is overcome by providing an optimized solution. Search results provide varying number of URLs for each search query. Processing a large no of URLs significantly reduces the response time of the search engine. Hence overload of the system must be avoided. The concept of Load Shedding is used for this purpose. Dropping data tuples during times of overload is known as Load Shedding. When Load Shedding concept is used, it is possible that a highly trustworthy URL is missed. Hence a trade-off occurs between reducing the response time and processing all resultant URLs. Thus an optimized solution has been proposed so that response time is also reduced and a maximum possible no of URLs are also processed.

## 4 ARCHITECTURE

The architecture presented below in Fig. 1 shows the overall architecture of Enhanced Trustworthy and High-Quality Information Retrieval System for Web Search Engines [1] with Load Shedder incorporated to handle overload conditions.

The user enters the search query and also selects WIQA Quality Policies in the web search engine user interface. The Searcher retrieves the corresponding search results from the database. The resultant URLs are given to the Load Shedder. The Load Shedder implements the Optimal Load Shedding Algorithm presented in section 5. The Load Shedder calls the Trust Evaluator which calculates the trustworthiness value for each of the retrieved URL. All the URLs are filtered based on the calculated trustworthiness values.

The filtered URLs are then given to the Quality Sub-System, which stores the URLs in the Named Graphs. The quality level for each URL is calculated based on three metrics – Content, Context and Ratings. These metrics are decided based on the policies initially selected by the user. The Decision Maker finally calculates the quality level for each URL by giving a weight factor to the three metrics. Finally the user receives Trustworthy and High-Quality information from the web search engine.

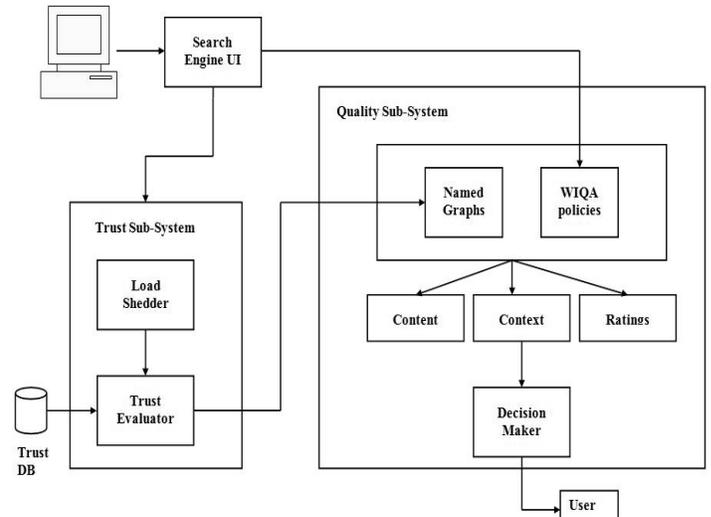

Fig. 1 Overall Architecture of High Performance Trustworthy Information Retrieval System

The architecture presented in Fig.4.2 shows the Load Shedder and its components.

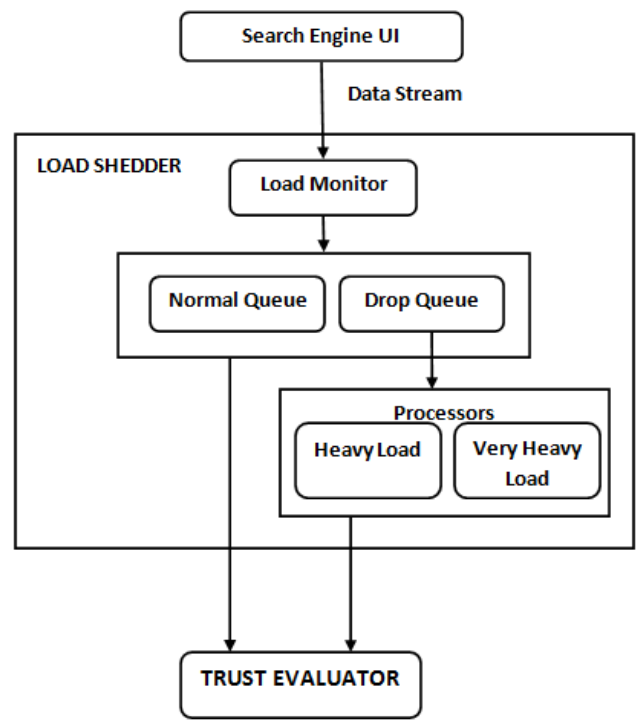

Fig. 2 Architecture of Load Shedder

The URLs obtained from the search results is the data stream. The entire load of URLs is given to the Load Monitor which monitors the load and decides upon the 3 parameters required for the algorithm.

The three parameters include
- Uload,
- Ucapacity and
- Uthreshold.

Uload represents the current load of URLs in the system. Ucapacity is the number of URLs which can be processed by the system within a certain response time which is the deadline of the system. Uthreshold is the number of URLs above Ucapacity that can be processed within an optimum response time selected for overload conditions.

The Optimal Load Shedding Algorithm can handle 3 kinds of load conditions to the system. They are Normal Load, Heavy Load and Very Heavy Load.

### 4.1 Normal Load Processing

Under the Normal Load condition, the number of URLs retrieved is not more than Ucapacity.

$$Uload <= Ucapacity$$

In this case, all the URLs are put in the Normal Queue. From the Normal Queue, the URLs are retrieved one by one and processed by the Trust Evaluator, which calculates the trustworthiness of the URL. The response time of the search engine in this case is well below the optimum level.

### 4.2 Heavy Load Processing

Under the Heavy Load condition, the number of URLs retrieved is more than Ucapacity but not more than Ucapacity + Uthreshold.

$$Uload <= Ucapacity + Uthreshold$$

In this case, the URLs up to Ucapacity are put in the Normal Queue and the remaining URLs are put in the Drop Queue. The Normal Queue is processed as mentioned in section 4.1. The URLs in the Drop Queue are initially checked with the Trust DB. All the URLs for which the trustworthiness value has already been stored are assigned with the value retrieved from the Trust DB and the corresponding URLs are removed from the Drop Queue.

Now the current time taken is checked with the deadline (optimum response time during overload conditions). If the deadline is not yet met, the remaining URLs in the Drop Queue are retrieved one by one and sent to the Trust Evaluator for trustworthiness calculation. Once the deadline is met, all the remaining URLs in the Drop Queue are assigned with an average trustworthiness value. Thus the deadline is met and all the URLs are also processed without dropping them from the system.

### 4.3 Very Heavy Load Processing

Under the Very Heavy Load condition, the number of URLs retrieved is more than Ucapacity + Uthreshold.

$$Uload > Ucapacity + Uthreshold$$

In this case, the URLs up to Ucapacity are put in Normal Queue and the remaining URLs are put in Drop Queue. The Normal Queue is processed as mentioned in section 4.1. In soft real time DSMS applications [6] missing deadline does not cause significant impact. So before starting the processing of Drop Queue, the deadline of the system is increased by a specific value. This value is calculated by giving a weight based on Uload and the optimum response time the user needs.

The URLs in the Drop Queue are checked against the Trust DB. All the URLs whose trustworthiness is already calculated and present in the Trust DB are removed from the Drop Queue. Until the modified deadline of the system is met, each of the remaining URLs from the Drop Queue is sent to the Trust Evaluator for processing and is removed from the queue. Once the modified deadline is met, an average trustworthiness value is given to all the remaining URLs in the Drop Queue. Thus, even during very heavy load conditions, all the URLs are processed without dropping and a near optimum response time is given to the user along with trustworthy information.

## 5 ALGORITHM

### 5.1 Load Shedder Algorithm

```
procedure Load_Shedder
begin
  Calculate Uload
  if (Uload <= Ucapacity) then
    normal_load()
  else if (Uload <= Ucapacity + Uthreshold) then
    heavy_load()
  else if (Uload > Ucapacity + Uthreshold) then
    vheavy_load()
  end if
end Load_Shedder
```

### 5.2 Normal Load Processing

```
procedure normal_load
begin
  while (URL in Normal Queue)
  begin
    if (URL present in DB) then
      assign trustworthiness
    else
      Calculate trustworthiness
    end if
  end while
end normal_load
```

### 5.3 Heavy Load Processing

```
procedure heavy_load
begin
  normal_load()
  if (current_time < deadline) then
  begin
    while (URL in Drop Queue)
    begin
      if (URL present in DB) then
        assign trustworthiness
        remove URL from Drop Queue
      end if
    end while
    while (URL in Drop Queue)
    begin
      if(current_time < deadline) then
        Calculate trustworthiness
        Remove URL from Drop Queue
      end if
    end while
  else
    Assign average trustworthiness for all URLs in
   Drop Queue
  end if
end heavy_load
```

### 5.4 Very Heavy Load Processing

```
procedure vheavy_load
begin
  Increase deadline
  heavy_load()
end vheavy_load
```

## 6 RESULT ANALYSIS

An open source search engine called NUTCH is used to get experimental results. The Existing System is [1] and the Proposed System is the architecture described in this paper.

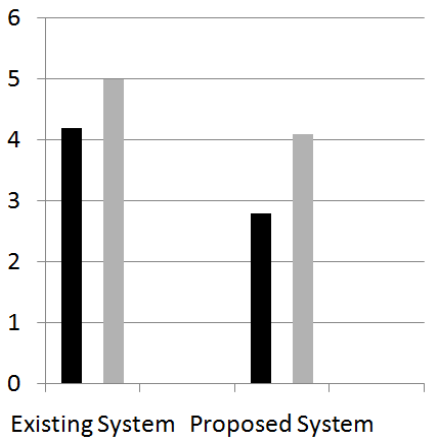

Fig.3.1 (a)

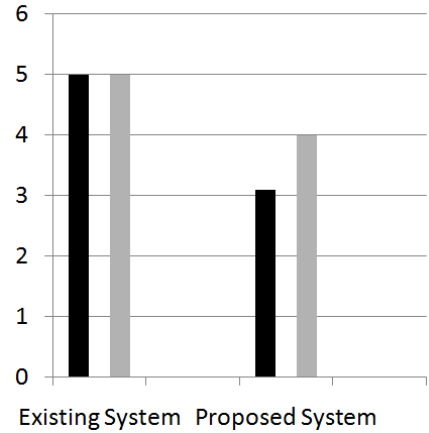

Fig.3.1 (b)

■ Response Time
■ Trustworthiness

Fig.3.1 Experimental Results for (a) Heavy Load (b) Very Heavy Load

Fig.3.1 (a) shows the Response Time and Trustworthiness of the Existing System and Proposed System under Heavy Load conditions. All the calculations are based on a scale of 5. The Trustworthiness is maintained at the maximum level and the Response Time ranges between 4 – 4.5. But, in the Proposed System the Load Shedding algorithm maintains an optimum level of Trustworthiness which is 4.1 and the Response Time is greatly reduced to 2.8.

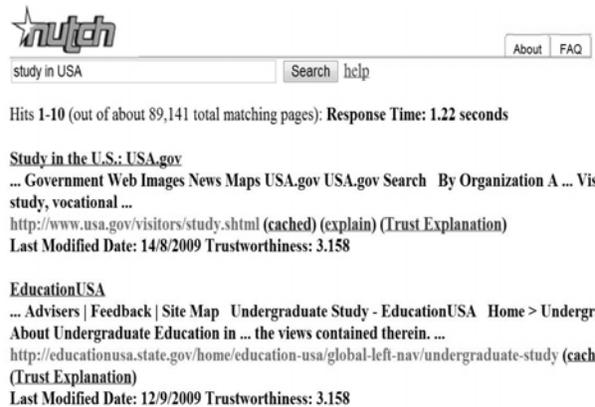

Fig.3.2 (a) Under HEAVY LOAD: Existing System

Fig.3.1 (b) shows the Response Time and trustworthiness of the Existing System and Proposed System under Very Heavy Load conditions. As in Fig.3.1 (a), the Response Time and Trustworthiness are at the maximum for the Existing System. The Proposed System provides a good Response Time of 3.1 even under very heavy load conditions. The level of Trustworthiness is also made optimum at 4.

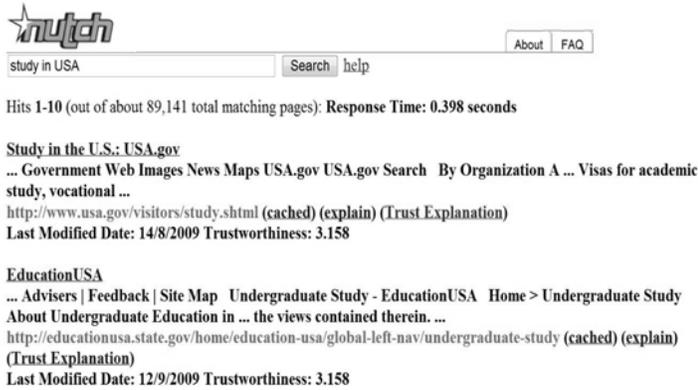

Fig.3.2 (b) Under HEAVY LOAD: Proposed System

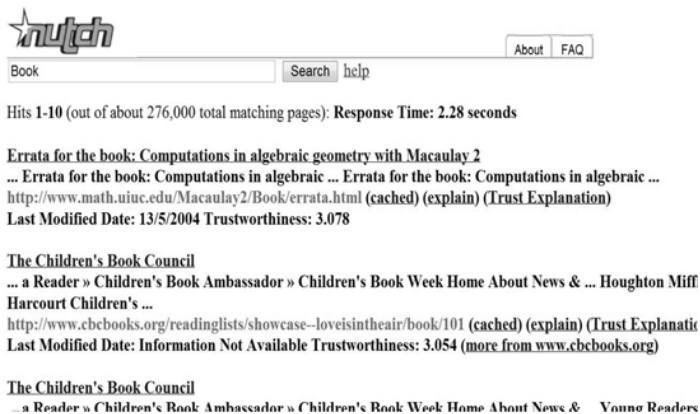

Fig.3.2 (c) Under VERY HEAVY LOAD: Existing System

Fig.3.2 (a), (b), (c) and (d) show the output of the Nutch search engine under Heavy Load and Very Heavy Load conditions. Fig.3.2 (a) shows the Existing System's response for the search query "Stud y in USA". The response time for this search query was found to be 1.22 seconds. Under same conditions and using the same database, the response time for the Proposed System for the same query was found to be 0.398 seconds. This is shown in Fig.3.2 (b).

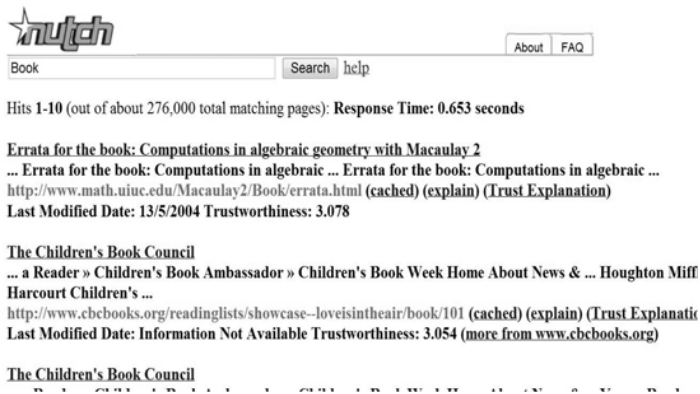

Fig.3.2. (d) Under VERY HEAVY LOAD: Proposed System

Fig.3.2 (c) shows the Existing System's response for the search query "book". Since this is a very common keyword, the number of search results retrieved is very large and the system is put under Very Heavy Load. The response time for this search query was found to be 2.28 seconds. Under same conditions and using the same database, the response time for the Proposed System for the same query was found to be 0.653 seconds.

Thus the response time for the search results is reduced even during Heavy and Very Heavy Load conditions, while using the Optimal Load Shedding Algorithm.

## 7 CONCLUSION AND FUTURE WORK

The user gets trustworthy and high-quality information from the search engine with good response time. Even under overloaded conditions when a bursty load is processed it is required to give trustworthy results in optimized response time. An optimal load shedding algorithm is proposed to handle this overload condition.

The proposed algorithm gives optimized response time, compromising the accuracy in calculating the trustworthiness under very heavy load conditions. For future work, to handle this very heavy overload condition an adaptive approach is analyzed to reduce this trade off.

**Sumalatha Ramachandran** received the B.E. degree in Computer Science from Madras University and M.E. degree in Electronics Engineering from Madras Institute of Technology, Anna University. She has completed her Ph.D in the area of information retrieval system. She is currently working in the area of Semantic Information retrieval using visualization techniques, Database management systems and web services.

**Sujaya Paulraj** is a member of the IEEE, IEEE Computer Society and student of Dept. Of Information Technology, Anna University.

**Sharon Joseph** is a member of ACM and student of Dept. Of Information Technology, Anna University.

**Vetriselvi Ramaraj** is a member of ACM and student of Dept. Of Information Technology, Anna University.